# The modified fundamental equations of quantum mechanics

Huai-Yu Wang
*Department of Physics, Tsinghua University, Beijing 100084, China*
wanghuaiyu@mail.tsinghua.edu.cn
794024763@qq.com phone number: +86-13717938266

**Abstract:** The Schrödinger equation, Klein–Gordon equation (KGE), and Dirac equation are believed to be the fundamental equations of quantum mechanics. Schrödinger's equation has a defect in that there are no negative kinetic energy (NKE) solutions. Dirac's equation has positive kinetic energy (PKE) and NKE branches. Both branches should have low-momentum, or nonrelativistic, approximations: one is the Schrödinger equation, and the other is the NKE Schrödinger equation. The KGE has two problems: it is an equation of the second time derivative so that the calculated density is not definitely positive, and it is not a Hamiltonian form. To overcome these problems, the equation should be revised as PKE- and NKE-decoupled KGEs. The fundamental equations of quantum mechanics after the modification have at least two merits. They are unitary in that all contain the first time derivative and are symmetric with respect to PKE and NKE. This reflects the symmetry of the PKE and NKE matters, as well as, in the author's opinion, the matter and dark matter of our universe. The problems of one-dimensional step potentials are resolved by utilizing the modified fundamental equations for a nonrelativistic particle.

**Résumé:** L'équation de Schrödinger, l'équation de Klein-Gordon (KGE) et l'équation de Dirac sont considérées les équations fondamentales de la mécanique quantique. L'équation de Schrödinger a un défaut en ce qu'il n'y a pas de solutions d'énergie cinétique négative (NKE). L'équation de Dirac a des branches d'énergie cinétique positive (PKE) et NKE. Les deux branches doivent avoir des approximations à faible impulsion, ou non relativistes: l'une est l'équation de Schrödinger et l'autre est l'équation NKE de Schrödinger. Le KGE a deux problèmes: c'est une équation de la dérivée seconde du temps de sorte que la densité calculée n'est pas définitivement positive, et ce n'est pas une forme hamiltonienne. Pour surmonter ces problèmes, l'équation doit être révisée en tant que KGE soit découplée de PKE et NKE. Les équations fondamentales de la mécanique quantique après la modification ont au moins deux mérites. Ils sont unitaires en ce sens qu'ils contiennent tous la première dérivée temporelle et sont symétriques par rapport à PKE et NKE. Cela reflète la symétrie des matières PKE et NKE, ainsi que, selon l'auteur, la matière et la matière noire de notre univers. Les problèmes de potentiels de pas unidimensionnels sont résolus en utilisant les équations fondamentales modifiées pour une particule non relativiste.

# I. INTRODUCTION

The fundamental equations of quantum mechanics (QM) refer to Schrödinger equation,[1-4] Klein-Gordon equation[5,6] (KGE) and Dirac equation,[7] which have been used up to now and are fully introduced in QM textbooks.[8,9] The latter two are relativistic quantum mechanics equations (RQMEs).

The Schrödinger equation[1-4] is the first of the QM equations (QMEs) and applies to microscopic particles with low-momentum motion. The "low-momentum motion" is the synonym of the "nonrelativistic motion" usually called. The name "nonrelativistic motion" usually means the motion described by Newtonian equations in classical mechanics and the motions described by Schrödinger equation in QM, and this term may be somehow misleading. The Newtonian equations determine the motion of bodies with low momentum which can be perfectly solved by special relativity, and Schrödinger equation determine the motion of particles with low momentum which can be perfectly solved by the RMQEs. The low-momentum motion is still a part of relativistic motion. Therefore, I will often use "low-momentum motion" and seldom mention "nonrelativistic motion".

Schrödinger tried to extend his equations to the case of relativity[4] but failed because the evaluated fine structure of the hydrogen atom was not consistent with the experimental one.[8] The Dirac equation explained these experiments almost perfectly.

In a previous work,[10] I became aware of the inconsistency between the Schrödinger equation and the low-momentum approximations of RQMEs and suggested how to remedy the inconsistency. The main result was that a relativistic particle had solutions of negative energy branch, which meant that a particle could have negative kinetic energy (NKE) besides positive kinetic energy (PKE). The NKE should still be retained when the particle performed low-momentum motion. In Ref. 10, an experiment was suggested to verify the existence of NKE electrons. The envisaged experiment used a scanning tunneling microscope (STM) device and let photons collide with the tunneling electrons in the region between the tip and sample surface. This would confirm that the electrons in the region are of NKE.

The work in Ref. 10 stimulates me to inspect the fundamental equations of QM as listed in Table I. The Dirac equation is no doubt correct but might not be used and understood correctly. An example was Klein's paradox.[11-23] This paradox also existed for KGE.[12,15,19,22]

As for KGE, there was another RQME describing the relativistic motion of spin-0, called the Salpeter equation.[24-41] The author thinks that there are inconsistencies between the fundamentals, so the equations need to be modified.

The author has the following five basic viewpoints.

The first is that whether the Schrödinger equation applies to potential barrier regions or not, where a particle's energy $E$ is less than potential $V$, has never been verified experimentally nor derived theoretically, which was mentioned previously.[10]

The second one is that, in both classical and quantum theories, a relativistic theory,

applicable to the motion of any momentum, is a complete one, while a nonrelativistic one, applicable to low-momentum motion, is not. From a nonrelativistic theory itself, one is unable to find its drawbacks. For example, nobody had found any problem in Newtonian mechanics itself. Only after Einstein had established the special relativity and its low-momentum approximation been taken, people knew that Newtonian mechanics was not valid for high-momentum motion. Similarly, it is difficult to find the defects of Schrödinger equation from itself. Only when taking low-momentum approximations from RQMEs can one know what is lost in Schrödinger equation.

The third viewpoint is that the PKE and NKE solutions of RQMEs should be treated on an equal footing. One example using this viewpoint is that Klein's paradox was solved perfectly.[42] The uniform procedure was applied to the cases of both spin-1/2 and spin-0 particles. The results were physically reasonable, and were applied to explain the very strong transmissivity of neutrinos and Dirac fermions in graphene.

The fourth viewpoint is that all the properties of RQMEs, e. g., the symmetry in respect to PKE and NKE, should be retained when nonrelativistic approximations are made.

This means that low-momentum QME ought to have the properties that RQMEs have. However, it was pointed out[10] that there were three properties that RQMEs had that were lost in the Schrödinger equation. The first was that an RQME had both PKE and NKE solutions; the second was that the probability currents of the PKE and NKE solutions were opposite in direction; the third was that when potential $V$ took the contrary sign, $V \to -V$, the eigenvalues $E$ were either, $E \to -E$.

The fifth viewpoint is that our observable substances are of PKE, and dark substances are of NKE. That is to say, in my opinion, dark substances are all that we have known; they are dark to us simply because they are of NKE. For instance, nobody has detected particles in the barrier regions. People detect particles only when they are out of potential barriers. Thus, our universe has matter-dark matter symmetry, i. e., PKE-NKE symmetry. In Ref. 43, the statistical mechanics and thermodynamics of NKE systems were presented, and the symmetry concerning PKE and NKE systems was revealed; see the tables there. The equations of motion of macroscopic NKE bodies were derived,[44] and again, PKE-NKE symmetry was exhibited. Later, more topics with respect to NKE matter will be studied. As long as the physical properties of NKE systems are explored, we will see PKE-NKE symmetry. Tables II and III in this paper below embody PKE-NKE symmetry.

It is generally believed that the NKE solutions of a particle's Dirac equation is that of its antiparticle with PKE. In Appendix A, we distinguish the NKE equation of a particle with charge $q$ and the PKE equation of its antiparticle with charge $-q$.

People may think that quantum field theory (QFT) was developed after the RQMEs had been established, which was to deal with problems in the field of elementary particles at a higher level. Indeed, when utilizing the RQMEs, people found some tricky problems, such as Klein's paradox, the difficulty of negative probability of Klein-Gordon equations, ect. Hence, the QFT was generated. I like to point out that the QFT did not solve these problems but merely circumvented them. The QFT deals with problems concerning particles' creation and annihilation. It cannot be used to solve underlying problems which do not involve particles' creation and annihilation. For examples, the QFT is unable to give answers of the following problems: whether Schrödinger equation is applicable in region where a particle's energy is less than potential is neither experimentally verified nor derived theoretically rigorously; Klein's paradox when a relativistic particle encounters a step potential; the difficulty of negative probability arising from Klein-Gordon equation; Schrödinger equation, as a low-momentum approximation of the RQMEs, fails to inherit the symmetry with respect to PKE and NKE solutions that the RQMEs have; the transition from the fundamental equation of QM to the equations of motion; ect.

I have solved the mentioned problems within the scope of RQMEs, as done in my previous work.[10,42,43,44] It is my belief that any problem can be figured out within the RQMEs, as long as particles' creation and annihilation are not concerned. In this view, I have not found any defect of Dirac equation.

In the course of solving the problems, I found that the fundamental equations of QM need to be made perfect. For example, in my previous work, NKE Schrödinger equation was proposed to remedy the defects embodied in Schrödinger equation. All the fundamental equations of QM ought to be logically consistent.

This paper manages to modify the fundamental equations of QM to become appropriate forms. In section 2, the defects of Schrödinger equation and KGE are presented. The equations after modification can get rid of the defects. The form of Salpeter equation prompted us how to modify the equations of relativistic particles with spin-0. The modified equations are uniformly of first time derivative and symmetric with respect to positive and negative energy branches. Section 3 gives an example indicating that the results calculated by the modified equations differ from those by using only Schrödinger equation. Section 4 is our conclusions.

## II. FUNDAMENTAL EQUATIONS IN SYMMETRIC FORMS

In Newtonian mechanics of classical physics, the kinetic energy of a body is defined by

$$K_{(+)} = \frac{\boldsymbol{p}^2}{2m}. \tag{1}$$

Here the subscript (+) refers to PKE. The momentum $\boldsymbol{p}$ is real. When a body is subject to a potential $V$, its energy is

$$E_{(+)} = \frac{\boldsymbol{p}^2}{2m} + V > V. \tag{2}$$

In special relativity, a body's energy is expressed by

$$E_{(+)} = \sqrt{m^2 c^4 + c^2 \boldsymbol{p}^2}. \tag{3}$$

When the body is subject to a potential $V$, its energy becomes

$$E_{(+)} = \sqrt{m^2c^4 + c^2\boldsymbol{p}^2} + V > V. \qquad (4)$$

When the momentum is very low, Eqs. (3) and (4) can be, respectively, approximated to (1) and (2) plus a static energy $mc^2$. Please note that in both Eqs. (2) and (4), it is impossible for a particle's energy to be $E_{(+)} < V$, which is a remarkable feature of classical mechanics.

Here, it should be noted that Eqs. (1) and (2) are only valid for low-momentum motion, while Eqs. (3) and (4) stand for any momentum value, i. e., the momentum can be from zero to an arbitrarily large number. When the momentum is very low, the particle's energy can be expressed by either Eq. (2) or (4) because, besides the constant static energy, they merely differ by a negligibly small quantity. Using (1) and (2) is simpler, but using (3) and (4) is more precise.

In QM, the particle's motion, when its momentum was very low, was believed to obey the Schrödinger equation.

$$i\hbar \frac{\partial}{\partial t}\psi = (-\frac{\hbar^2}{2m}\nabla^2 + V)\psi. \qquad (5)$$

Then, KGE was proposed to describe the relativistic motion of spin-0 particles.

$$(i\hbar \frac{\partial}{\partial t} - V)^2 \Psi = (m^2c^4 - c^2\hbar^2\nabla^2)\Psi. \qquad (6)$$

Shortly afterwards, Dirac equation was developed, which was for the relativistic motion of spin-1/2 particles.

$$i\hbar \frac{\partial}{\partial t}\Psi = (-ic\hbar\boldsymbol{\alpha} \cdot \nabla + mc^2\beta + V)\Psi. \qquad (7)$$

Table I. Fundamental equations of QM being used, where $\boldsymbol{p} = -i\hbar\nabla$.

| | | | |
|---|---|---|---|
| Relativistic motion | Spin-1/2 particles | Dirac equation $i\hbar\frac{\partial}{\partial t}\Psi = (c\boldsymbol{\alpha} \cdot \boldsymbol{p} + mc^2\beta + V)\Psi$ | |
| | Spin-0 particle | Klein-Gordon equation $(i\hbar\frac{\partial}{\partial t} - V)^2\psi = (m^2c^4 + c^2\boldsymbol{p}^2)\psi$ | Problems: second time derivative, negative probability, not a Hamiltonian form. |
| Nonrelativistic motion | | Schrödinger equation $i\hbar\frac{\partial}{\partial t}\psi = (\frac{1}{2m}\boldsymbol{p}^2 + V)\psi$ | Problems: lack of negative energy branch, there is no solution when $V \to -V$. |

Equations (5), (6) and (7) are fundamental equations of QM being used nowadays, and are listed in Table I. In Table I, some problems related to Schrödinger equation and KGE are briefly illustrated. Here we have a detailed analysis of these problems.

First, the Dirac equation is inspected. From it, the energy of a free relativistic particle is expressed by

$$E_{(\pm)} = \pm\sqrt{m^2c^4 + c^2\mathbf{p}^2}. \tag{8}$$

There are two energy branches. The positive branch $E_{(+)}$ is exactly the same as (3), i. e., it has classical correspondence.

The negative branch does not have classical correspondence. Its explicit form is

$$E_{(-)} = -\sqrt{m^2c^4 + c^2\mathbf{p}^2}. \tag{9}$$

This is just the contrary number of (3). Thus, since (3) contains positive static energy and PKE, naturally, (9) contains negative static energy and NKE. When the particle is subject to a potential $V$, then the positive branch becomes Eq. (4) and the negative branch becomes

$$E_{(-)} = -\sqrt{m^2c^4 + c^2\mathbf{p}^2} + V < V. \tag{10}$$

When the momentum is very low, Eq. (10) can be approximated, with the static energy dropped, to be

$$E_{(-)} = -\frac{\mathbf{p}^2}{2m} + V = K_{(-)} + V < V, \tag{11}$$

where it is denoted that

$$K_{(-)} = -\frac{\mathbf{p}^2}{2m}. \tag{12}$$

Here the NKE of low momentum is in the form of (12), which is rigorously derived from (9) by low-momentum approximation, and the momentum $\mathbf{p}$ is real. Please note that, in both Eqs. (10) and (11), the energy $E_{(-)}$ is always less than potential $V$. This feature comes from QM. Conversely, if a particle's energy is larger than its potential, the energy should be expressed by Eqs. (2) or (4), and if a particle's energy is less than its potential, the energy should be expressed by Eqs. (10) or (11).

When one takes transformation

$$\Psi = \psi_{(+)}e^{-imc^2t/\hbar} \tag{13}$$

in the Dirac equation, then the wave function $\psi_{(+)}$ satisfies Schrödinger equation (5) in low-momentum approximation. That is to say, the Schrödinger equation (5) is a low-momentum approximation of Dirac equation.

Now let us consider such a situation: a free particle is moving with very low momentum. On the one hand, its motion obeys the Schrödinger equation, from which the particle's energy is solved in Eq. (2). On the other hand, the Dirac equation is precisely applicable, from which the particle's energy has two branches, as indicated by Eq. (8). Thus, in the Schrödinger equation, the negative energy branch is absent.

The author believes that the Dirac equation is correct, and the negative energy branch should be treated on an equal footing as the positive one. Thereby, the negative branch should be retained even in low-momentum motion. Fortunately, this can be achieved by the transformation

$$\Psi = \psi_{(-)} e^{imc^2 t/\hbar}. \tag{14}$$

By this transformation and taking low-momentum approximation, we obtain from the Dirac equation that

$$i\hbar \frac{\partial}{\partial t} \psi_{(-)} = (\frac{\hbar^2}{2m} \nabla^2 + V)\psi_{(-)}. \tag{15}$$

From Eq. (15), a free particle has energy (12), which is of NKE. Therefore, Eq. (15) is called NKE Schrödinger equation. This equation, as one low-momentum approximation of the Dirac equation, embodies the negative energy branch and applies to the cases of $E_{(+)} < V$, while the Schrödinger equation applies to the cases of $E_{(+)} > V$. The combination of the two equations inherits the properties of the Dirac equation in low-momentum motion and is listed in the bottom row in Table II.

Table II. Modified fundamental equations of QM, where $H_0 = \sqrt{m^2 c^4 - c^2 \hbar^2 \nabla^2}$.

| | | In PKE region, $E > V$ | In NKE region, $E < V$ |
|---|---|---|---|
| Relativistic motion | Spin-1/2 particle | Dirac equation $i\hbar \frac{\partial}{\partial t} \Psi = (-ic\hbar \boldsymbol{\alpha} \cdot \nabla + mc^2 \beta + V)\Psi$ | |
| | Spin-0 particle | PKE-decoupled KGE (Salpeter equation) $i\hbar \frac{\partial}{\partial t} \psi_{(+)} = (H_0 + V)\psi_{(+)}$ | NKE-decoupled KGE $i\hbar \frac{\partial}{\partial t} \psi_{(-)} = (-H_0 + V)\psi_{(-)}$ |
| Low-momentum motion | | Schrödinger equation $i\hbar \frac{\partial}{\partial t} \psi_{(+)} = (-\frac{\hbar^2}{2m} \nabla^2 + V)\psi_{(+)}$ | NKE Schrödinger equation $i\hbar \frac{\partial}{\partial t} \psi_{(-)} = (\frac{\hbar^2}{2m} \nabla^2 + V)\psi_{(-)}$ |

It is emphasized that the low-momentum equations and the Dirac equation in Table II are in the form of

$$i\hbar \frac{\partial \psi}{\partial t} = H\psi .\tag{16}$$

where $H$ is the Hamiltonian. That is to say, the Hamiltonian of a system is definitely known.

Now, we turn to KGE (6). It has severe problems coming from the fact that Eq. (6) contains the second time derivatives. Let us consider a spin-0 particle performing very low-momentum motion. It should obey the Schrödinger equation, which contains the first time derivative. Suppose that we can accelerate the particle in some way until it performs relativistic motion. Then, it should obey KGE. At this point, we would like to ask a question: in the course of the increase of the particle's momentum, at which momentum does the equation that the particle satisfies transits from the first time derivative one to the second time derivative one, and why? We are unable to give a satisfactory answer.

Because KGE contains the second time derivative, to solve the wave function, the initial conditions needed are the initial value of the wave function:

$$\psi(\boldsymbol{r}, t=0) \tag{17}$$

and the initial value of the first time derivative of the wave function:

$$[\frac{\partial}{\partial t}\psi(\boldsymbol{r},t)]_{t=0} . \tag{18}$$

In contrast, in solving the Schrödinger equation, only condition (17) is needed.

A contradiction arises when a spin-0 particle performs low-momentum motion. On the one hand, its wave functions can be solved by the Schrödinger equation under the initial condition (17). On the other hand, it can be solved precisely from KGE but with an additional initial condition (18): is the condition (18) necessary or not on earth?

A consequence of the second time derivative of KGE was that the probability density derived from this equation could be negative, the famous negative probability difficulty.[8,9]

This problem demonstrates that either Eq. (5) or (6) are questionable. It is believed that the Schrödinger equation correctly describes the motion of particles for low momenta and energies $E > V$, and it is a low-momentum approximation of the Dirac equation.

One more problem is that the Klein–Gordon equation (6) is not of the form of (16), which means we do not know what the Hamiltonian is, and strictly speaking, there is no concept of eigenenergies. This is eccentric. Feshbach and Villars[22,45] reformed Klein–Gordon equation (6) into the form of (16). Their procedure was to separate the wave function into two parts

$$\psi = \varphi + \chi , \tag{19}$$

and to set the two parts as two components of a spinor. In this way the Klein-Gordon equation was reformed to be Feshbach-Villars equation.

$$i\hbar \frac{\partial}{\partial t}\begin{pmatrix}\varphi\\\chi\end{pmatrix}=H\begin{pmatrix}\varphi\\\chi\end{pmatrix}. \tag{20}$$

Thus, there is a Hamiltonian in Eq. (20). This Hamiltonian $H$ is not Hermitian. In a non-Hermitian system, the norms of eigen wave functions vary with time so that they cannot always be normalized,[22] and the eigen wave functions belonging to different eigenvalues may not be orthogonal to each other.[22,46] To implement evaluation, a generalized inner product was defined[22] for the system described by Eq. (20), which was different from the inner product usually used in QM. By the way, we think that the wave function of a spin-0 particle should be a scalar instead of a spinor.

The present author tried to obtain the classical limits from the KGE and Feshbach–Villars equation, but failed. I give my reasoning in appendix B.

Thus, it is believed that the KGE needs to be modified. Let us determine the proper equations for relativistic particles with spin-0.

In my previous papers,[10,42] I pointed out that when the potential $V$ was a piecewise constant, KGE could be factorized as

$$(i\hbar\frac{\partial}{\partial t}+H_0+V)(i\hbar\frac{\partial}{\partial t}-H_0+V)\psi=0, \tag{21}$$

where

$$H_0=\sqrt{m^2c^4-c^2\hbar^2\nabla^2}. \tag{22}$$

The order of the two parentheses in Eq. (21) can be exchanged. Thus, we obtained two decoupled equations.

$$i\hbar\frac{\partial}{\partial t}\psi_{(+)}=(\sqrt{m^2c^4-c^2\hbar^2\nabla^2}+V)\psi_{(+)} \tag{23}$$

and

$$i\hbar\frac{\partial}{\partial t}\psi_{(-)}=(-\sqrt{m^2c^4-c^2\hbar^2\nabla^2}+V)\psi_{(-)}. \tag{24}$$

Obviously, the energy of a particle satisfying Eq. (23) is related to Eq. (4), and that satisfying (24) is related to (10). Therefore, Eq. (23) is called PKE-decoupled KGE, and (24) is called NKE-decoupled KGE. Using these forms, we solved Klein's paradox for spin-0 particles.[42]

Equations (23) and (24) were achieved when the potential was a piecewise constant. I extend them now by allowing the potential to be arbitrary and think that they are the appropriate equations that relativistic particles with spin-0 should obey. From now on, the potential in Eqs. (23) and (24) can be arbitrary. Equation (23) was, in fact, suggested before[24-26] and called Salpeter equation. There have been investigations of the Salpeter equation with various potentials.[27-41] Yndurain proved that Eq. (23) was of relativistic invariance,[24] as a relativistic equation should be. Equation (24) is the NKE counterpart of (23)

Equations (23) and (24) offer the following merits. First, their low-momentum approximations are naturally the Schrödinger equation and NKE Schrödinger equation, respectively. Explicitly, by use of Eq. (13) in (23) and by use of (14) in (24), we can obtain Eqs. (5) and (15) after taking low-momentum approximations. These processes are the same as those of deriving Eqs. (5) and (15) from Dirac equation. Alternatively, one can also expand the square root of $H_0$ (22), make a low-momentum approximation and drop the constant terms $\pm mc^2$, to obtain Eqs. (5) and (15). Second, the forms of Eqs. (23) and (24) lead to correct expressions of probability density, avoiding the difficulty of negative probability.[10] Third, these two equations helped to correctly calculate the reflection coefficients of a relativistic particle with spin-0 through the one-dimensional step potentials.[24] The achieved reflection curves were qualitatively the same as those of a Dirac particle. Besides, it is believed that the decoupled KGEs will have more usage. For instance, the Wigner function of the Salpeter equation was explored.[47]

Equations (23) and (24) are listed in Table II which thus gathers all the modified fundamental QMEs. The equations listed in Table II have the first time derivative uniformly so that all can be written in the form of Eq. (16).

In present QM textbooks, when dealing with particles' low-momentum motion, all the problems are solved by the Schrödinger equation. According to the author's opinion, the problems concerning potential barriers where $E < V$ should be revisited. In Section III below, an example is given.

Microscopic particles with PKE can, using interactions between them, compose macroscopic PKE bodies that observe classical mechanics. In Ref. 44, the author obtained Newtonian and relativistic mechanics from the Schrödinger equation and PKE-decoupled KGE under macroscopic approximation.

Table III. Laws of motion for PKE and NKE matters. RM refers to relativistic motion.

| Matter | Laws of motion | | | PKE matter | NKE matter |
|---|---|---|---|---|---|
| Macro-scopic bodies | Classical mechanics | Low-momentum motion | | Newtonian mechanics for PKE bodies | Newtonian mechanics for NKE bodies |
| | | RM | | Relativistic mechanics for PKE bodies | Relativistic mechanics for NKE bodies |
| Micro-scopic particles | Quantum mechanics | Low-momentum motion | | Schrödinger equation | NKE Schrödinger equation |
| | | RM | Spin-0 particles | Salpeter equation | NKE decoupled KGE |
| | | | Spin-1/2 particles | Dirac equation | |

Naturally, microscopic NKE particles can also, by interactions between them,

compose macroscopic bodies. In Ref. 44, the author obtained the equations of motion for NKE bodies from the NKE Schrödinger equation and PKE-decoupled KGE under macroscopic approximation. They are called, respectively, Newtonian mechanics and relativistic mechanics for NKE bodies, as listed in Table III here. In that work, the macroscopic approximation was taken. After that, I was aware that the forms of decoupled KGEs (23) and (24) above could stand for microscopic particles.

Our universe is of good symmetry with respect to PKE and NKE matters. Tables II and III demonstrate the symmetry of fundamental equations of motion. The author's opinion is that the Schrödinger equation, PKE-decoupled KGE, and the PKE solutions of the Dirac equation describe the motion of PKE matter we have studied so far, while the NKE Schrödinger equation, NKE-decoupled KGE, and NKE solutions of the Dirac equation describe the motion of dark matter, which is of NKE. Personally, NKE matter is a synonym of dark matter. In a previous work,[43] the fundamental formalism of the statistical mechanics and thermodynamics for NKE systems was given. In statistical mechanics and thermodynamics, there is also good symmetry with respect to PKE and NKE systems.

In the next section, I will provide an example to disclose the discrepancy between the results obtained by low-momentum equations in Tables I and II.

## III. RE-CALCULATION OF REFLECTION COEFFICIENT OF A PARTICLE BY A ONE-DIMENSIONAL STEP POTENTIAL

### A. An infinitely wide potential

The potential is

$$V(x) = \begin{cases} 0, & x \leq 0 \\ V_0, & x > 0 \end{cases}. \tag{25}$$

Please see Fig. 1(a). A particle with energy $E$ and momentum $q$ is incident from the left and moves rightwards. Let us calculate the reflection coefficient.

At first, we retrospect what was done in usual QM textbooks. Actually, the Schrödinger equation in Table I was used.

In region I, $x < 0$, the particle's energy and its wave function are, respectively:

$$E = q^2/2m \tag{26}$$

and

$$\psi_I = e^{iqx/\hbar} + Be^{-iqx/\hbar}. \tag{27}$$

The two terms in (27) are, respectively, incident and reflective waves, with $B$ being reflection amplitude. In region II, $x > 0$, its energy and wave function are believed to be

$$E = p^2/2m + V_0 \tag{28}$$

and

$$\psi_{II} = Fe^{ipx/\hbar}, \tag{29}$$

with $F$ being transmission amplitude. At the boundary $x=0$, the wave function should be continuous and smooth. These conditions lead to

$$1 + B = F \tag{30a}$$

and

$$q(1-B) = pF. \tag{30b}$$

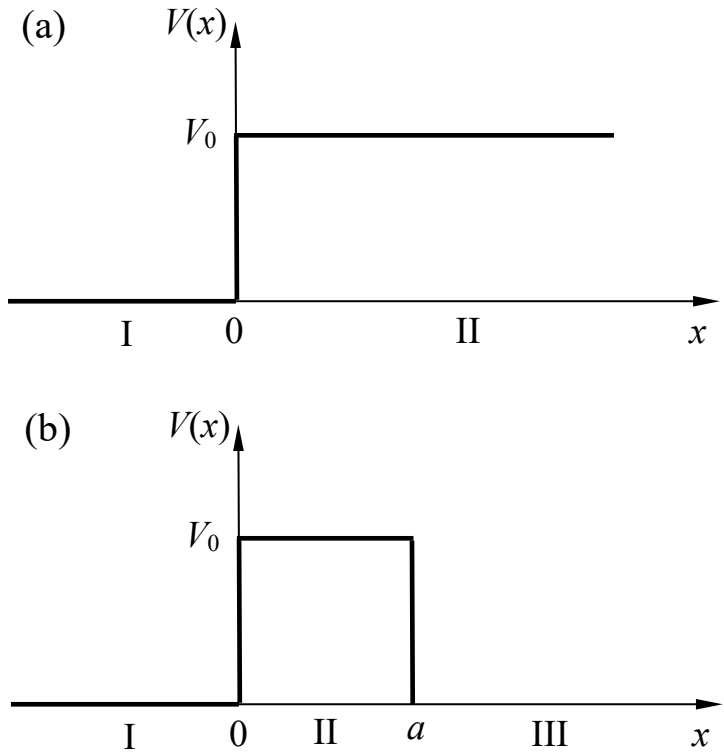

FIG. 1. One-dimensional potential barrier with height $V_0$. (a) Infinitely wide. (b) Finitely wide with width $a$.

From Eqs. (30) the reflection amplitude is obtained:

$$B = \frac{q-p}{q+p}. \tag{31}$$

When $E > V_0$, $p$ is real and reflection coefficient is

$$R = |B|^2 = \left(\frac{V_0}{2E - V_0 + 2\sqrt{E(E-V_0)}}\right)^2. \tag{32}$$

Numerical results are depicted by the dashed line in Fig. 2(a).

As $E < V_0$, it seemed from (28) that

$$p = i\mu \tag{33}$$

was an imaginary number, and hence the reflection coefficient calculated by (32) was

$$R = 1, \tag{34}$$

i. e., the particle totally reflected. This was because in the region $x > 0$ the wave function was thought decaying exponentially. This conclusion seemed plausible, but was questionable.

As a matter of fact, Table II tells us that in the case of $E < V_0$, NKE Schrödinger equation (15) should be employed, so that Eq. (28) should be replaced by (11).

Hence in region II, $x > 0$, the equation should be

$$\frac{\hbar^2}{2m}\frac{d^2}{dx^2}\psi_{II} + V_0\psi_{II} = E\psi_{II}. \tag{35}$$

The wave function is

$$\psi_{II} = Fe^{ipx/\hbar}. \tag{36}$$

It is a plane wave with the same form as (29). Correspondingly, the energy-momentum relation of the particle is

$$E = -p^2/2m + V_0. \tag{37}$$

Thus, $p$ is real, and Eq. (33) is questionable. Please compare Eqs. (28) and (37). They are manifestation of Eqs. (2) and (11). In Eq. (37), $E - V_0 = -p^2/2m$ is the NKE. Energy is always kinetic energy plus potential energy, no matter whether the kinetic energy is positive or negative. Please also note that in both (29) and (36), the momentum $p$ is real.

In region $x < 0$, Eqs. (26) and (27) remain valid.

We now have simultaneous equations (27) and (36) plus energy–momentum relations (26) and (37). The boundary conditions have exactly the same forms as (30). The resultant reflection coefficient is

$$R = |B|^2 = \left(\frac{V_0 - 2E}{V_0 + 2\sqrt{E(V_0 - E)}}\right)^2. \tag{38}$$

Numerical results are plotted in Fig. 2(a) by solid line. The discussion of the curves in Fig. 2(a) is in Subsection III.C.

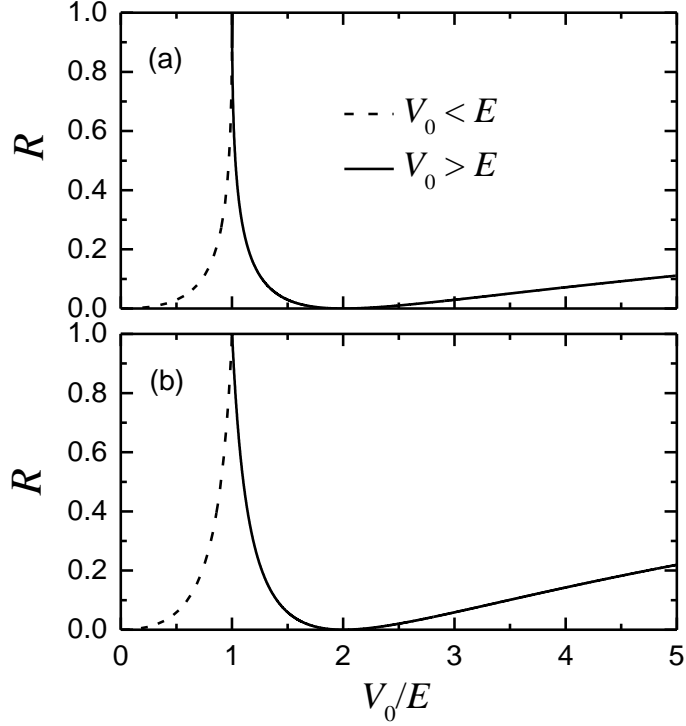

FIG. 2. Reflection coefficient $R=|B|^2$. (a) The potential barrier is FIG. 1(a). Dashed line is calculated by Eq. (32) and solid line is calculated by Eq. (38). (b) The potential barrier is FIG. 1(b), and $\cot(pa/\hbar) = 1/2$ is taken. Dashed line is calculated by Eq. (50) and solid line is calculated by Eq. (56).

**B. A finitely wide potential**

The potential is sketched in Fig. 1(b).

$$V(x) = \begin{cases} 0, & x \leq 0 \\ V_0, & 0 < x \leq a \\ 0, & x > a \end{cases}. \tag{39}$$

A particle with energy $E$ and momentum $q$ is incident from $-\infty$ and moves rightwards.

We first recall what has been completed in QM textbooks.

In regions I and III, the wave functions are, respectively, written in the following forms.

$$\psi_\mathrm{I} = e^{iqx/\hbar} + Be^{-iqx/\hbar}, \quad x < 0. \tag{40}$$

$$\psi_\mathrm{III} = Ge^{iqx/\hbar}, \quad x > a. \tag{41}$$

The energy-momentum relation is

$$E = q^2/2m. \tag{42}$$

In region II, one has to distinguish the cases of $V_0 < E$ and $V_0 > E$.

1. $V_0 < E$

In region II, the wave function can be written as

$$\psi_{II} = F_1 e^{ipx/\hbar} + F_2 e^{-ipx/\hbar}, \quad 0 < x < a. \tag{43}$$

The energy-momentum relation is

$$E = p^2/2m + V_0. \tag{44}$$

The three functions (40), (41) and (43) should meet the conditions that at boundaries $x = 0$ and $x = a$ the wave function and its derivative are continuous. Then, the following four equations are obtained.

$$1 + B = F_1 + F_2. \tag{45}$$

$$F_1 e^{ipa/\hbar} + F_2 e^{-ipa/\hbar} = G e^{iqa/\hbar}. \tag{46}$$

$$q(1 - B) = p(F_1 - F_2). \tag{47}$$

$$p(F_1 e^{ipa/\hbar} - F_2 e^{-ipa/\hbar}) = qG e^{iqa/\hbar}. \tag{48}$$

From these four equations and with the help of Eqs. (42) and (44), the reflection amplitude is solved.

$$B = \frac{V_0}{2E - V_0 + 2i\sqrt{E(E - V_0)} \cot(pa/\hbar)}. \tag{49}$$

The reflection coefficient $R = |B|^2$ is

$$R = \frac{V_0^2}{(2E - V_0)^2 + 4E(E - V_0)\cot^2(pa/\hbar)}. \tag{50}$$

Numerical results are potted by dashed line in Fig. 2(b).

From Eq. (50), it is seen that as the barrier height is fixed, $R$ oscillates with the barrier width $a$.

As $pa/\hbar = n\pi$, $R = 0$. \tag{51}

This is resonant transmission. In this case, from Eqs. (45)-(48) the coefficients are solved:

$$B = 0, G = 1, F_1 = \frac{1}{2}(1 + \frac{q}{p}), F_2 = \frac{1}{2}(1 - \frac{q}{p}). \tag{52}$$

The reflection maxima are as follows.

$$\text{As } pa/\hbar = (n+\frac{1}{2})\pi, \quad R = \frac{V_0^2}{(2E-V_0)^2}. \tag{53}$$

2. $V_0 > E$

In this case, according to QM textbooks, the energy-momentum relation is still Eq. (44), from which the momentum has to be an imaginary number, $p = ik$. By the boundary conditions at $x = 0$ and $x = a$, the reflection coefficient is solved to be

$$R = \frac{V_0^2}{(2E-V_0)^2 + 4E(V_0 - E)\coth^2(ka/\hbar)}. \tag{54}$$

As $a \to \infty$, $R \to 1$. This is the conclusion in QM textbooks.

However, in the viewpoint of present work, in region II, the energy-momentum relation should be

$$E = -p^2/2m + V_0, \tag{55}$$

and the wave function is still of the form of Eq. (44). Therefore, in this case, we have wave functions Eqs. (40), (41) and (43) available, but the energy-momentum relationships should be Eqs. (42) and (54). By the boundary conditions at $x = 0$ and $x = a$, the reflection coefficient is solved to be

$$R = \frac{(2E-V_0)^2}{V_0^2 + 4E(V_0 - E)\cot^2(pa/\hbar)}. \tag{56}$$

Numerical results are plotted by the solid line in Fig. 2(b). The discussion of curves in Fig. 2(b) is in the next subsection.

Under fixed barrier height, the reflection coefficient oscillates with the barrier width, which is the same behavior as (50). The condition of resonant transmission is also (51), and under these conditions, the coefficients are the same as (52)

$$\text{As } pa/\hbar = (n+\frac{1}{2})\pi, \quad R = (1 - \frac{2E}{V_0})^2. \tag{57}$$

This is the reflection maximum.

### C. Discussions

In the two panels of Fig. 2 that the reflection coefficient curves have the same profiles, and they have following common features.

As $V_0 = 0$, there is no barrier so that $R = 0$. As $V_0$ increases starting from zero to $E$, reflection coefficient $R$ rises from 0 to 1. At $V_0 = E$, $p = 0$. That is to say, the particle cannot move in the barrier, so that it reflects totally.

In the range $0<V_0<2E$, there is a symmetry: the exchange $V_0-E \Leftrightarrow E-V_0$ makes Eq. (38) become (32) and vice versa, and makes Eq. (50) become (45) and vice versa. It is reflected in Fig. 2 that within $0<V_0/E<2$, the curves are symmetric with respect to $V_0/E=1$. As $V_0=2E$, it is easily derived from (26) and (37), as well as from (42) and (44), that $p=q$. That is to say, the value of the particle's momentum does not alter when it passes through the potential wall, as if there is no force acting on it. This situation is equivalent to that it is not scattered, and is similar to the case of "complete impedance matching" in electromagnetic materials. Therefore, the transmission coefficient is 1.

As $V_0>2E$, $R$ again rises with $V_0$ monotonically. As $V_0 \to \infty$, $R \to 1$. Facing the infinitely high barrier, the particle has to reflect totally.

In a previous paper,[42] I have solved the same potential problems for relativistic particles by means of the equations listed in Table II here. Now I compare the curves there with those in Figs. 2 here. The qualitative behaviors are the same. There is always a transmission alley at $V_0=2E$. For a finitely wide potential barrier, the resonant transmission conditions are also (41).

In all cases as $V_0>2E$, $R$ rises from zero monotonically. For a Dirac particle, when $V_0 \to \infty$, roughly speaking, $R \to m^2c^4/E^2 <1$. That is to say, even facing an infinitely high potential, there is still some probability for the particle to transmit, and the greater the particle's energy or the less its static mass, the greater the transmission probability, which might provide an insight why neutrinos could transmit through almost everywhere.[42]

The discrepancies between relativistic and low-momentum particles when $E<V$ are that the energy of the former has a gap while that of the latter have not.

For a relativistic particle, when the energy-momentum relationship meets $c^2p^2=(E-V)^2+m^2c^4<0$, there is an energy gap in the range $[-mc^2+V, mc^2+V]$ with a width $2mc^2$. Within this gap, the momentum $p$ is an imaginary number. Suppose that the height of the step potential barrier is within $-mc^2+E<V<mc^2+E$. The wave functions in the barrier are of exponential forms. For an infinitely wide potential, the reflection coefficient curves within this range is 1, called reflection platform; for a finitely wide potential, the curve in this range is less than 1.[42]

For a low-momentum particle, the energy has no gap, see Eqs. (28), (37), (44) and (54). There is no possibility for the momentum $p$ to be imaginary. The curves within $-mc^2 + E < V_0 < mc^2 + E$ for a relativistic particle contract to a point at $V_0 = E$ in Figs. 2(a) and 2(b).

The usual treatment of the tunneling problem in QM textbooks gives (54), which is questionable, since the NKE Schrödinger equation is not touched. Nevertheless, nobody has noticed that because no superficial contradiction appeared. As for relativistic motion, neglecting the negative energy solution caused the reflection coefficient to be larger than 1, so Klein's paradox emerged.

Here, we emphasize that in the regions $E < V$, the NKE Schrödinger equation is not generated out of nothing but derived from the Dirac equation rigorously by means of transformation (14) and then low-momentum approximation.

## IV. CONCLUSIONS

The unmodified fundamental equations of quantum mechanics (QM) are the Dirac equation, Klein–Gordon equation (KGE), and Schrödinger equation. After some defects are pointed out, they equations are modified as follows. The Dirac equation for spin-1/2 particles remains unchanged, which has positive kinetic energy (PKE) and negative kinetic energy (NKE) branches. A relativistic particle with spin-0 observes PKE- (NKE-) decoupled KGE for $E > V$ ( $E < V$ ). A low-momentum particle observes the Schrödinger equation (NKE Schrödinger equation) for $E > V$ ( $E < V$ ).

These equations uniformly have the first time derivative. The forms of the equations embody the symmetry with respect to PKE and NKE. In this way, the NKE and PKE branches are treated on an equal footing. The modified equations remedy the logical inconsistency of the unmodified equations.

The reflection coefficient of a low-momentum particle encountering a one-dimensional step potential with either infinite or finite width is calculated by means of the Schrödinger equation and NKE Schrödinger equation. The results are physically reasonable and have similar behavior to those of a relativistic particle.

This work is supported by the National Key Research and Development Program of China (2018YFB0704304-3).

# APPENDIX A DISTINGUISHING THE NKE EQUATION OF A PARTICLE WITH CHARGE $q$ AND THE PKE EQUATION OF AN ANTIPARTICLE WITH CHARGE $-q$

It is usually thought that the NKE solutions of the Dirac equation represent antiparticles. In this appendix, the author stresses that only one equation, instead of a pair of equations, is dealt with.

A particle with spin-1/2 and electric charge $q$ observes Dirac equation,

$$i\hbar \frac{\partial}{\partial t}\Psi^{(P)} = H(q)\Psi^{(P)}, \tag{A1}$$

where

$$H(q) = c\boldsymbol{\alpha}\cdot(-i\hbar\nabla - q\boldsymbol{A}) + mc^2\beta + q\phi. \tag{A1a}$$

Here a superscript (P) labels the particle. This equation has both PKE and NKE energies, denoted by $E_{(+)}^{(P)}$ and $E_{(-)}^{(P)} = -E_{(+)}^{(P)}$, respectively. The corresponding equations for stationary states are, respectively,

$$H(q)\Psi_{(+)}^{(P)} = E_{(+)}^{(P)}\Psi_{(+)}^{(P)} \tag{A2a}$$

and

$$H(q)\Psi_{(-)}^{(P)} = E_{(-)}^{(P)}\Psi_{(-)}^{(P)} = -E_{(+)}^{(P)}\Psi_{(-)}^{(P)}. \tag{A2b}$$

Here I want to stress that the solutions of (A2b) is the NKE ones, not that of an antiparticle. The charge in (A1) is already assigned as $q$, not $-q$. Equation (A1) is for a particle with $q$, not for a pair of particles with $q$ and $-q$.

Its antiparticle carries charge $-q$, and observes the following Dirac equation:

$$i\hbar \frac{\partial}{\partial t}\Psi^{(A)} = H(-q)\Psi^{(A)}, \tag{A3}$$

where $H(-q)$ is Eq. (A1a) with $q$ being substituted by $-q$. Now the superscript (A) labels the antiparticle. This equation has also both PKE and NKE energies, denoted by $E_{(+)}^{(A)}$ and $E_{(-)}^{(A)} = -E_{(+)}^{(A)}$, respectively. The corresponding equations of stationary states are, respectively,

$$H(-q)\Psi_{(+)}^{(A)} = E_{(+)}^{(A)}\Psi_{(+)}^{(A)} \tag{A4a}$$

and

$$H(-q)\Psi_{(-)}^{(A)} = E_{(-)}^{(A)}\Psi_{(-)}^{(A)}. \tag{A4b}$$

The antiparticle with charge $-q$ itself has both the PKE and NKE solutions. The NKE solutions of (A4b) belongs to the antiparticle with $-q$ instead of particle $q$.

**What I concentrate on is the pair of equations (A2a) and (A2b), both being for a particle with charge $q$.**

I do not talk about the equations (A2b) and (A4a), because they belong to two particles. I discuss equations for only one particle.

The charge conjugation operator changes the equation of a particle with charge $q$ into that of the antiparticle with charge $-q$, i. e., change one equation into another. I merely talk about the equations that are not acted by the charge conjugation operator.

Now, let us take low-momentum approximations.

First, we do this for Eq. (A1). We put down the wave function in the form of a two-component spinor.

$$\Psi^{(P)} = \begin{pmatrix} \varphi^{(P)} \\ \chi^{(P)} \end{pmatrix}. \tag{A5}$$

Let

$$\Psi^{(P)} \to \Psi^{(P)} e^{-imc^2 t/\hbar}. \tag{A6}$$

By this transformation, Eq. (A1) is recast to be the equations that the two components satisfy.

$$i\hbar \frac{\partial \varphi^{(P)}}{\partial t} = c\boldsymbol{\sigma} \cdot (\boldsymbol{p} - q\boldsymbol{A})\chi^{(P)} + q\phi\varphi^{(P)}. \tag{A7a}$$

$$i\hbar \frac{\partial \chi^{(P)}}{\partial t} = c\boldsymbol{\sigma} \cdot (\boldsymbol{p} - q\boldsymbol{A})\varphi^{(P)} + q\phi\chi^{(P)} - 2mc^2 \chi^{(P)}. \tag{A7b}$$

It is seen that the $\varphi^{(P)}$ is the larger component and the $\chi^{(P)}$ is the smaller one. From (A7) it is approximated that

$$\chi^{(P)} \approx \frac{1}{2mc} \boldsymbol{\sigma} \cdot (\boldsymbol{p} - q\boldsymbol{A})\varphi^{(P)}. \tag{A8}$$

This is substituted into (A7) to get

$$i\hbar \frac{\partial \varphi^{(P)}}{\partial t} = \frac{1}{2m} [\boldsymbol{\sigma} \cdot (\boldsymbol{p} - q\boldsymbol{A})]^2 \varphi^{(P)} + q\phi\varphi^{(P)}. \tag{A9}$$

This is the Schrödinger equation for a particle with charge $q$. This low-momentum approximation is usually down in QM textbooks.

Now we take another transformation for Eq. (A1) which has not been used in QM textbooks. Let

$$\Psi^{(P)} \to \Psi^{(P)} e^{imc^2 t/\hbar}. \tag{A10}$$

It differs from (A6) only in the sign in the exponent. By this transformation, Eq. (A1) becomes the equations that the two components satisfy.

$$i\hbar \frac{\partial \varphi^{(P)}}{\partial t} = c\boldsymbol{\sigma} \cdot (\boldsymbol{p} - q\boldsymbol{A}) \chi^{(P)} + q\phi \varphi^{(P)} + 2mc^2 \varphi^{(P)}. \tag{A11a}$$

$$i\hbar \frac{\partial \chi^{(P)}}{\partial t} = c\boldsymbol{\sigma} \cdot (\boldsymbol{p} - q\boldsymbol{A}) \varphi^{(P)} + q\phi \chi^{(P)}. \tag{A11b}$$

This time, the $\varphi^{(P)}$ is the smaller component and the $\chi^{(P)}$ is the larger one. From (A11a) it is approximated that

$$\varphi^{(P)} \approx -\frac{1}{2mc} \boldsymbol{\sigma} \cdot (\boldsymbol{p} - q\boldsymbol{A}) \chi^{(P)}. \tag{A12}$$

This is substituted into (A11b) to get

$$i\hbar \frac{\partial \chi^{(P)}}{\partial t} = -\frac{1}{2m} [\boldsymbol{\sigma} \cdot (\boldsymbol{p} - q\boldsymbol{A})]^2 \chi^{(P)} + q\phi \chi^{(P)}. \tag{A13}$$

This is the NKE Schrödinger equation for the particle with charge $q$.

For the equation of the antiparticle with charge $-q$, Eq. (A3), we can also do the low-momentum approximation. Again, the wave function in (A3) is put down in the form of a two-component spinor.

$$\Psi^{(A)} = \begin{pmatrix} \varphi^{(A)} \\ \chi^{(A)} \end{pmatrix}. \tag{A14}$$

Let

$$\Psi^{(A)} \to \Psi^{(A)} e^{-imc^2 t/\hbar}. \tag{A15}$$

By this transformation, Eq. (A3) is recast to be the following two equations.

$$i\hbar \frac{\partial \varphi^{(A)}}{\partial t} = c\boldsymbol{\sigma} \cdot (\boldsymbol{p} + q\boldsymbol{A}) \chi^{(A)} - q\phi \varphi^{(A)}. \tag{A16a}$$

$$i\hbar \frac{\partial \chi^{(A)}}{\partial t} = c\boldsymbol{\sigma} \cdot (\boldsymbol{p} + q\boldsymbol{A}) \varphi^{(A)} - q\phi \chi^{(A)} - 2mc^2 \chi^{(A)}. \tag{A16b}$$

The $\varphi^{(A)}$ is larger and the $\chi^{(A)}$ is smaller. From (A16b) it is approximated that

$$\chi^{(A)} \approx \frac{1}{2mc} \boldsymbol{\sigma} \cdot (\boldsymbol{p} + q\boldsymbol{A}) \varphi^{(A)}. \tag{A17}$$

This is substituted into (A16a) to get

$$i\hbar \frac{\partial \varphi^{(A)}}{\partial t} = \frac{1}{2m}[\boldsymbol{\sigma}\cdot(\boldsymbol{p}+q\boldsymbol{A})]^2 \varphi^{(A)} - q\phi\varphi^{(A)}. \tag{A18}$$

This is the Schrödinger equation for the antiparticle with charge $-q$.

Another transformation for Eq. (A3) is that

$$\Psi^{(A)} \to \Psi^{(A)} e^{imc^2 t/\hbar}. \tag{A19}$$

After this transformation, Eq. (A3) becomes the following two equations.

$$i\hbar \frac{\partial \varphi^{(A)}}{\partial t} = c\boldsymbol{\sigma}\cdot(\boldsymbol{p}+q\boldsymbol{A})\chi^{(A)} - q\phi\varphi^{(A)} + 2mc^2 \varphi^{(A)}. \tag{A20a}$$

$$i\hbar \frac{\partial \chi^{(A)}}{\partial t} = c\boldsymbol{\sigma}\cdot(\boldsymbol{p}+q\boldsymbol{A})\varphi^{(A)} - q\phi\chi^{(A)}. \tag{A20b}$$

This time, the $\varphi^{(A)}$ is smaller and the $\chi^{(A)}$ is larger. From (A20a) it is approximated that

$$\varphi^{(A)} \approx -\frac{1}{2mc}\boldsymbol{\sigma}\cdot(\boldsymbol{p}+q\boldsymbol{A})\chi^{(A)}. \tag{A21}$$

This is substituted into (A20b) to get

$$i\hbar \frac{\partial \chi^{(A)}}{\partial t} = -\frac{1}{2m}[\boldsymbol{\sigma}\cdot(\boldsymbol{p}+q\boldsymbol{A})]^2 \chi^{(A)} - q\phi\chi^{(A)}. \tag{A22}$$

This is the NKE Schrödinger equation for the antiparticle with charge $-q$.

It is stressed again that the Eqs. (A2a) and (A2b) are the equations with the PKE and NKE solutions. The low-momentum approximation of the former is Eq. (A9) and that of the latter is Eq. (A13). These equations are all for a particle with charge $q$.

For the antiparticle with charge $-q$, the equations with the PKE and NKE solutions are Eqs. (A4a) and (A4b), and their low-momentum forms are, respectively, Eqs. (A18) and (A22).

Dirac equations for a particle and its antiparticle and their low-momentum approximations are listed in Table IV.

Table IV. Dirac equations for a particle and its antiparticle and their low-momentum approximations. $h_{(\pm)}(q) = \pm\frac{1}{2m}(\boldsymbol{p} - q\boldsymbol{A})^2 + q\phi$, $h_{(\pm)}(-q) = \pm\frac{1}{2m}(\boldsymbol{p} + q\boldsymbol{A})^2 - q\phi$.

| Particle | Electric charge $q$ | Electric charge $-q$ |
|---|---|---|
| Dirac equation | $i\hbar\frac{\partial}{\partial t}\Psi^{(P)} = H(q)\Psi^{(P)}$ | $i\hbar\frac{\partial}{\partial t}\Psi^{(A)} = H(-q)\Psi^{(A)}$ |
| Schrödinger equation after transformation $\Psi = \psi_{(+)}e^{-imc^2 t/\hbar}$ | $i\hbar\frac{\partial}{\partial t}\psi_{(+)}^{(P)} = h_{(+)}(q)\psi_{(+)}^{(P)}$ | $i\hbar\frac{\partial}{\partial t}\psi_{(+)}^{(A)} = h_{(+)}(-q)\psi_{(+)}^{(A)}$ |
| NKE Schrödinger equation after transformation $\Psi = \psi_{(-)}e^{imc^2 t/\hbar}$ | $i\hbar\frac{\partial}{\partial t}\psi_{(-)}^{(P)} = h_{(-)}(q)\psi_{(-)}^{(P)}$ | $i\hbar\frac{\partial}{\partial t}\psi_{(-)}^{(A)} = h_{(-)}(-q)\psi_{(-)}^{(A)}$ |

**APPENDIX B KLEIN-GORDON EQUATION AND FESHBACH-VILLARS EQUATION CANNOT BE USED TO DERIVE EQUATION OF MOTION IN CLASSICAL MECHANICS**

Bohm believed long before[48-55] that the Schrödinger equation with $H$,

$$i\hbar\frac{\partial \psi}{\partial t} = H\psi, \tag{B1}$$

could transit to equation of Newtonian mechanics. The procedure is as follows, but I narrate it in my own way.

First, we construct two equations

$$\frac{\partial \rho}{\partial t} = \psi^*\frac{\partial \psi}{\partial t} + \psi\frac{\partial \psi^*}{\partial t} = \frac{1}{i\hbar}(\psi^* H\psi - \psi H^*\psi^*) \tag{B2}$$

and

$$i\hbar(\psi^*\frac{\partial \psi}{\partial t} - \psi\frac{\partial \psi^*}{\partial t}) = \psi^* H\psi + \psi H^*\psi^*. \tag{B3}$$

The particle's density probability is defined as

$$\rho = \psi^*\psi. \tag{B4}$$

The wave function is put down in the form of

$$\psi = Re^{iS/\hbar}, \tag{B5}$$

where both $R$ and $S$ are real numbers. This was called the hydrodynamic model by Bohm. In this model, both $R$ and $S$ have specific physical meanings: the square of $R$ is just the density probability

$$\rho = R^2, \tag{B6}$$

and $S$ is the action of the system, the significance of which is clearly shown in Eqs. (B8)-(B10) below.

We now substitute the expression of the wave function (B5) into (B1), or alternatively, into Eqs. (B2) and (B3), and then take the limit $\hbar \to 0$. This approximation eliminates the quantum effect and is called hydrodynamic approximation or the hydrodynamic limit. Through this procedure, one obtains

$$\frac{\partial \rho}{\partial t} + \nabla \cdot \boldsymbol{j} = 0. \tag{B7}$$

from (B2) and

$$-\frac{\partial S}{\partial t} = K + V \tag{B8}$$

from (B3).[56]

In Eqs. (B7) and (B8), there is no Planck constant and no operator, so that they are believed to be equations for macroscopic bodies, i.e., the equations in classical mechanics. Now, all the quantities in these two equations are for macroscopic bodies.

Equation (B7) is apparently a continuity equation. The expression of $\boldsymbol{j}$ can be gained through Eq. (B2).

Equation (B8) is the Hamilton–Jacoby equation in classical mechanics.

Equations (B7) and (B8) had been derived long ago, but nobody really obtained the Newtonian equation of motion until the present author's previous work.[44] The method was to illustrate the standard formula in classical mechanics:

$$-\frac{\partial S}{\partial t} = H. \tag{B9}$$

Here $H$ is the Hamiltonian of the classical mechanics' system. It is kinetic energy plus potential. Momentum is evaluated by

$$\boldsymbol{p} = \nabla S. \tag{B10}$$

Both kinetic and potential energies are functions of coordinates $\boldsymbol{r}$ and momenta $\boldsymbol{p}$. One of Hamilton formulas is

$$\boldsymbol{v} = \dot{\boldsymbol{r}} = \frac{\partial H}{\partial \boldsymbol{p}}. \tag{B11}$$

Here, a dot on the top of a quantity represents taking its time derivative. By means of Legendre transformation,

$$L = \boldsymbol{p} \cdot \dot{\boldsymbol{r}} - H, \tag{B12}$$

Lagrangian was obtained. The Lagrangian, in turn, was put into Euler-Lagrange

equation

$$\frac{d}{dt}\frac{\partial L}{\partial \dot{r}} - \frac{\partial L}{\partial r} = 0 \tag{B13}$$

to achieve the equation of motion in classical mechanics. Along this routine, the transition from QM equations to the equations of classical mechanics was implemented.

The Schrödinger equation, NKE Schrödinger equation, and PKE- and NKE-decoupled KGEs are in the form of (B1) as listed in Table II.

In my previous work,[44] I performed the following tasks.

Starting from the Schrödinger equation and carrying out the procedure (B1)-(B13), Newtonian equations were obtained, which described the motion of macroscopic PKE bodies.

Starting from the NKE Schrödinger equation and carrying out the procedure (B1)-(B13), the equations describing the motion of macroscopic PKE bodies, which were thought dark macroscopic bodies.

Starting from PKE-decoupled KGE and carrying out the procedure (B1)-(B13), equations of motion in special relativity were obtained, which described the motion of PKE relativistic bodies.

Starting from NKE-decoupled KGE and carrying out the procedure (B1)-(B13), equations describing the motion of NKE relativistic bodies were obtained.

In all of the four cases, the inclusion of electromagnetic potentials was considered. The systems containing more than one body were considered. A PKE (NKE) system produced positive (negative) pressure. These conclusions were in agreement with those in Ref. 43.

In that work, the author gave the reason that it was impossible to obtain equations of motion in classical mechanics from the Dirac equation. The main reason was that spin has no classical correspondence.

The author thinks that if the KGE and Feshbach–Villars equation, are correct QM equations, they should also be used to derive the equations of motion in classical mechanics through the procedure (B1)-(B13). Below, I argue that they cannot be so.

The Klein–Gordon equation, when there is a time-independent electromagnetic field, is expressed as

$$(i\hbar\frac{\partial}{\partial t} - q\phi)^2 \psi = [c^2(\boldsymbol{p} - q\boldsymbol{A})^2 + m^2c^4]\psi. \tag{B14}$$

I let the wave function be the form of Eq. (B5) and substitute it into (B14). After some manipulation, I obtain the imaginary part identity:

$$\rho\frac{\partial^2 S}{\partial t^2} + (\frac{\partial S}{\partial t} + q\phi)\frac{\partial \rho}{\partial t} = c^2 \nabla \cdot (\rho\nabla S - q\boldsymbol{A}\rho). \tag{B15}$$

This is not an expected continuity equation. Extra ingredients appear due to the second time derivative in KGE. Therefore, I think that KGE is not a proper equation.

Now we turn to check the Feshbach–Villars equations. Let the wave function in (B14) be separated into two parts as follows.

$$\psi = \varphi + \chi, (i\hbar \frac{\partial}{\partial t} - q\phi)\psi = mc^2(\varphi - \chi). \tag{B16}$$

Then the new wave functions are expressed by

$$\varphi = \frac{1}{2}(\psi + \frac{1}{mc^2}(i\hbar \frac{\partial}{\partial t} - q\phi)\psi) \tag{B17a}$$

and

$$\chi = \frac{1}{2}(\psi - \frac{1}{mc^2}(i\hbar \frac{\partial}{\partial t} - q\phi)\psi). \tag{B17b}$$

Using (B14), the equations that $\varphi$ and $\chi$ satisfy are

$$i\hbar \frac{\partial \varphi}{\partial t} = \frac{1}{2m}(\boldsymbol{p} - q\boldsymbol{A})^2(\varphi + \chi) + (q\phi + mc^2)\varphi \tag{B18a}$$

and

$$i\hbar \frac{\partial \chi}{\partial t} = -\frac{1}{2m}(\boldsymbol{p} - q\boldsymbol{A})^2(\varphi + \chi) + (q\phi - mc^2)\chi. \tag{B18b}$$

These are Feshbach–Villars equations. To implement the transformation (B5), it is natural to transform the two new wave functions as follows.

$$\varphi = R_1 e^{iS/\hbar}, \chi = R_2 e^{iS/\hbar}, R_1 + R_2 = R. \tag{B19}$$

When Eq. (B19) is substituted into (B18), I obtain ill equations. In fact, we can sum the two equations of (B17), which is just Eq. (B16). Substituting (B5) and (B19) into (B16) leads to

$$i\hbar \frac{\partial R}{\partial t} - R \frac{\partial S}{\partial t} = mc^2(R_1 - R_2) + q\phi R. \tag{B20}$$

From this equation, no useful information can be extracted. Therefore, Feshbach–Villars equations failed to derive an equation of motion of classical mechanics, although, as QM equations, they are expected to have this function.